\documentclass{kapproc} 

\usepackage{procps} 

\usepackage[dvips]{graphicx}

\upperandlowercase

\kluwerbib 

\begin{document}

\articletitle[Simulating the high-z universe]
{Simulating the high-redshift Universe\\
in the sub-mm}

\author{Eelco van Kampen}

\affil{Institute for Astrophysics, University of Innsbruck\\
Technikerstr. 25, A-6020 Innsbruck\\
Austria
}
\email{Eelco.v.Kampen@uibk.ac.at}


\anxx{van Kampen\, Eelco}

\begin{abstract}
I present various simulations of an on-going large sub-mm survey, {\sl SHADES},
showing how constraints can be put on galaxy formation models and cosmology
from this survey.
\end{abstract}


\section{Introduction}

An important problem with most current galaxy formation models is how to establish
whether a set of model parameters that produces a good match to observations
is unique, as there are likely to be degeneracies amongst the various free
parameters of the model.
As most useful observational data used to constrain the model parameters
are obtained from our local universe, this `uniqueness problem' can be resolved
by comparing model predictions and observations at high redshift, which
in many respects is independent from a comparison at low redshift. 
Specifically, the {\sl SCUBA} half-degree extra-galactic survey
({\sl SHADES} for short; see {\tt http://www.roe.ac.uk/ifa/shades} for details)
will provide highly valuable observational data for this purpose.

\section{A new wide-area sub-mm survey: {\sl SHADES}}

{\sl SHADES} is a major new extragalactic survey with {\sl SCUBA}, 
the ``Submillimetre Common-User Bolometer Array'' (Holland et al.\ 1999),
covering 0.5 sq. degrees to a 4$\sigma$ detection limit of $S_{850}$ = 8\,mJy.
The data from {\sl SCUBA} will be supplemented with data from {\sl BLAST},
a ``Balloon-borne Large-Aperture Sub-millimeter Telescope'' (Devlin 2001),
which will undertake a series of nested extragalactic surveys at 250, 350 and 
500\,$\mu$m.

This wide-area survey will yield a substantial 
($\simeq200$) sample of bright unconfused sub-mm sources with meaningful 
redshift estimates ($\delta z \simeq \pm 0.5$).
The survey is designed to answer three fundamental questions.
What is the cosmic history of massive dust-enshrouded star-formation 
activity ? Are SCUBA sources the progenitors of present-day massive ellipticals ? 
What fraction of SCUBA sources harbour a dust-obscured AGN ? 

The crude but near-complete redshift information provided by {\sl BLAST}
is sufficient to answer the second question provided the survey covers
sufficient area and contains enough sources  to measure the clustering
of bright sub-mm sources on scales up to $\simeq$10\,Mpc.

\section{Modelling the high-redshift sub-mm population}

The assumption made here is that the bright sub-mm sources seen
by {\sl SCUBA} are dust-enshrouded starburst galaxies, as most show
no direct evidence for AGN activity (Almaini et al. 2003).
Simulating {\sl SHADES} is thus best done using a phenomenological
galaxy formation model, from which lightcones are constructed for all
galaxies with an 850\,$\mu$m flux over 8\,mJy.

The merging history of galaxy haloes is taken directly
from N-body simulations, which use special techniques to prevent
galaxy-scale haloes undergoing `overmerging' owing to inadequate
numerical resolution (van Kampen 1995). When haloes 
merge, a criterion based on dynamical friction is used to decide how
many galaxies exist in the newly merged halo. The most massive
of those galaxies becomes the single central galaxy to which
gas can cool, while the others become its satellites.

When a halo first forms, it is assumed to have
an isothermal-sphere density profile. A fraction
$\Omega_b/\Omega$ of this is in the form of gas
at the virial temperature, which can cool to form
stars within a single galaxy at the centre of the halo.
Energy output from supernovae reheats some of the
cooled gas back to the hot phase. 
Each halo maintains an internal account of the
amounts of gas being transferred between the two
phases, consumed by the formation of stars, and lost to
the environment.

\begin{figure}[t]
{\includegraphics[angle=270,width=\textwidth]{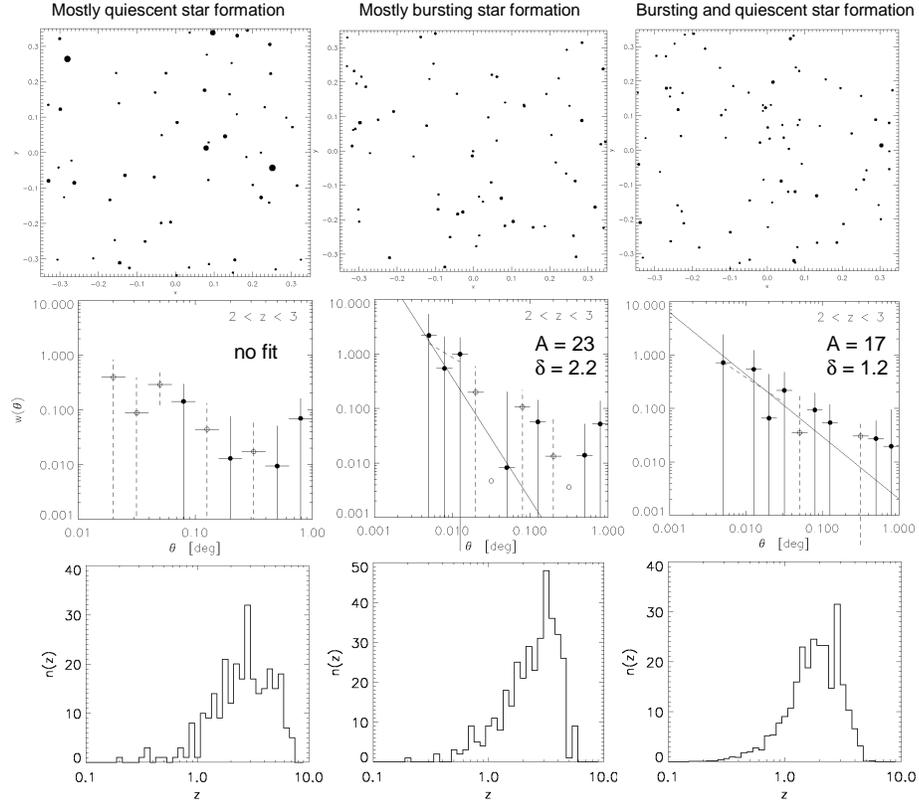}}
\caption{The three columns of panels represent results from three different
galaxy formation models, which mainly differ in how stars form (as indicated
at the top). The top row shows a single realization of a mock {\sl SHADES}, with the
size of the dots corresponding to the flux of each source. The middle row shows the
angular correlation functions for these maps, whereas the bottom row displays
the redshift distribution of the sources. Differences in these predictions
can be clearly observed.}
\end{figure}

The model includes two modes of star formation: quiescent
star formation in disks, and starbursts during merger events.
The evolution of the metals is followed, because the cooling of the
hot gas depends on metal content, and a stellar population of high
metallicity will be much redder than a low metallicity one of the
same age. It is taken as established that the population of brown dwarfs
makes a negligible contribution to the total stellar mass density,
and the model does not allow an adjustable $M/L$ ratio for the
stellar population.

The 850\,$\mu$m flux is assumed to be directly proportional
to the star formation rate, with an 8\,mJy flux corresponding to
1000 solar masses per year. The relation is taken to be fuzzy, so
that a random amount of flux (amounting to about 25 per cent) is
added/subtracted to reflect differences in dust temperatures and
grain sizes which are not yet modelled.

\section{Predictions for three different models}

Using the phenomenological galaxy formation model described above,
three different mock sub-mm surveys resembling {\sl SHADES} were
produced. These models only differ in their star formation laws,
in that one model has most stars forming quiescently in disks, one
has most stars forming in merger-induced starbursts, and the remaining
model has a mix of both. The resulting maps and predictions for the
angular clustering and redshift distribution are shown in Fig.\ 1.

While the number counts are similar (not shown), the clustering
properties, and to a lesser extent the redshift distributions,
are clearly different for these realizations. It seems that the
clustering strength depends on which star formation mode dominates,
with the burst model yielding relatively strong clustering, and
the quiescent model showing very little clustering. It has to be noted
that there is a fairly large spread in predicted clustering strengths
for different realizations of the {\it same model}, mainly due to
cosmic variance. Still, the trend remains visible after
considering 50 realizations for each model. The way forward is to
make optimal use of combining the difference in the clustering
{\it and} the redshift distribution. This work is currently in progress
(van Kampen et al. 2003).

\section{Conclusions and discussion}

The on-going large sub-mm survey {\sl SHADES} has the potential to
put significant constraints on galaxy formation models, and help resolve
the uniqueness problem of such models due to the uncertainties in
the assumptions, approximations, and choice of parameters. A potential
problem is that of cosmic variance: even though {\sl SHADES} is a
the largest extragalactic survey ever undertaken in the sub-mm waveband,
the total sky coverage is still much smaller than typically achieved
in the optical wavebands. However, two factors work in our favour:
the availability of (crude) redshift estimates for each of the sources,
and the expectation that clustering of bright sub-mm galaxies is
relatively strong (e.g. Percival et al. 2003, Scott et al. 2003, 
Webb et al. 2003, van Kampen et al. 2003).

\begin{acknowledgments}
This work was partly supported by the Austrian Science Foundation FWF
under grant P15868, and a PPARC rolling grant.
\end{acknowledgments}


\begin{chapthebibliography}{1}

\bibitem{omar}
Almaini, O., et al., 2003, MNRAS, 338, 303

\bibitem{devlin}
Devlin, M., 2001, astro-ph/001232

\bibitem{holland}
Holland, W., et al., 1999, MNRAS, 303, 659

\bibitem{evk95}
van Kampen, E., 1995, MNRAS, 273, 295

\bibitem{evkprep}
van Kampen, E., et al., 2003, in preparation

\bibitem{will}
Percival, W.J., Scott D., Peacock J.A., Dunlop J.S., 2003,
MNRAS, 338, 31

\bibitem{scott}
Scott, S.E., et al., 2003, in preparation

\bibitem{webb}
Webb, T.M., et al., 2003, ApJ, 587, 41

\end{chapthebibliography}
\end{document}